\documentclass[a4paper,11pt]{article}
\pdfoutput=1
\usepackage[english]{babel}  
\usepackage[tbtags]{amsmath}   
\usepackage{amsfonts}         
\usepackage{amssymb}           
\usepackage{graphicx}         
\usepackage[T1]{fontenc}   
\usepackage{ae}     
\usepackage{typearea}	       
\usepackage{scrpage2}          
\usepackage{lastpage}        
\usepackage[margin=10pt,font=small,labelfont=bf]{caption} 
\usepackage{dsfont}
\usepackage[plainpages=false]{hyperref}
\usepackage{multicol}
\usepackage{color}
\usepackage[bottom]{footmisc}
\usepackage{array}
\usepackage{epsfig}
\usepackage{graphicx}
\usepackage{pstricks}
\usepackage{pst-coil}
\usepackage[numbers,sort&compress]{natbib}
\usepackage{ulem}
\usepackage{color}
\usepackage{listings}

\typearea{18}                 

\hypersetup{
        unicode = true,
        pdftitle = {Neutrino Mass from a d=7 Effective Operator in an SU(5) SUSY-GUT Framework}, 
        pdfauthor = {M. B. Krauss, D. Meloni, W. Porod, W. Winter},
        colorlinks = true,
        linkcolor = blue,
        citecolor = blue,
        filecolor = blue,
        urlcolor = blue,
}

\newcommand{\T}{\mathsf{T}}

\newcommand{\ii}{\mathrm{i}}

\newcommand{\unit}[1]{\,\text{#1}}

\newcommand{\VEV}[1]{\langle #1 \rangle}

\newcommand{\med}[1]{$\mathbf{#1}$}

\newcommand{\vergiss}[1]{}

\newcommand{\be}{\begin{equation}}
\newcommand{\ee}{\end{equation}}
\newcommand{\bea}{\begin{eqnarray}}
\newcommand{\eea}{\end{eqnarray}}
\newcommand{\bi}{\begin{itemize}}
\newcommand{\ei}{\end{itemize}}

\newcommand{\ie}{{\it i.e.}}

\newcommand{\eg}{{\it e.g.}}

\newcommand{\cf}{{\it cf.}}

\newcommand{\eq}{Eq.}

\newcommand{\fig}{Fig.}

\newcommand{\Ref}{Ref.}
\newcommand{\Refs}{Refs.}

\newcommand{\Tab}{table}

\newcommand{\equ}[1]{\eq~(\ref{equ:#1})}
\newcommand{\figu}[1]{\fig~\ref{fig:#1}}

\font\TeXmanualfont=manfnt
\newcommand{\Vorsicht}{{\TeXmanualfont\char"7F\hskip 0.3em}}
  {\begin{list}{}{\leftmargin1em}%
     \item[\hbox to 0pt{\hss\Vorsicht}]}
  {\end{list}}

\begin{document}
\hspace{13cm} RM3-TH/12-21\\
\begin{center}
 \large \bfseries \sffamily 
Neutrino Mass from a d=7 Effective Operator in a SUSY-GUT Framework
\mdseries \rmfamily \medskip
 \normalsize
 
\medskip
 {\large
                Martin~B.~Krauss\footnote[1]{\makebox[1.cm]{Email:}
                martin.krauss@physik.uni-wuerzburg.de},
 		Davide~Meloni\footnote[2]{\makebox[1.cm]{Email:}
                meloni@fis.uniroma3.it}, 
 		Werner~Porod\footnote[3]{\makebox[1.cm]{Email:}
                porod@physik.uni-wuerzburg.de}, and
                Walter~Winter\footnote[4]{\makebox[1.cm]{Email:}
                winter@physik.uni-wuerzburg.de}
                }

\medskip

{\it

\footnotemark[1]${}^,$\footnotemark[3]${}^,$\footnotemark[4]
       Institut f{\"u}r Theoretische Physik und Astrophysik, Universit{\"a}t W{\"u}rzburg, \\
       Am Hubland, 97074 W{\"u}rzburg, Germany 

\medskip
       
\footnotemark[2]
	Dipartimento di Matematica e Fisica, Universit\`a degli Studi Roma Tre, \\Via del
la Vasca Navale 84, 00146 Roma, Italy 

} 

\medskip
\medskip
\medskip

 \today
\end{center}

\medskip

\begin{abstract}
Models, where neutrino mass originates from physics at the TeV scale and
 which are potentially 
testable at the LHC, need additional suppression mechanisms to describe the 
smallness of neutrino masses.  
We consider models in which neutrino mass is generated from the $d=7$ 
operator $LLH_uH_uH_dH_u$ in the context of SUSY-GUTs containing an SU(5) subgroup, where the $d=5$ Weinberg 
operator can be forbidden by a discrete symmetry.
That is, we identify the embeddings in GUT multiplets and their consequences for 
phenomenology and renormalization group evolution. We use a specific 
example to exemplify the challenges. In this case, additional heavy
$d$-quarks are predicted, which are constrained by cosmology, in particular,
by big bang nucleosynthesis and direct searches for heavy nuclei. 
We show that in 
the NMSSM extension of the model, the discrete symmetry needs to be broken, which can be the origin of deviations from tri-bimaximal 
mixings. Finally we demonstrate that our example is the only tree level decomposition
which is consistent with
perturbativity up to the GUT scale and neutrino mass generation by a leading $d=7$ 
contribution.

\end{abstract}

\bigskip

\section{Introduction}

The recent measurement of $\theta_{13}$ by Daya Bay \cite{An:2012eh} and RENO
\cite{Ahn:2012nd} has revealed that the last
of the three angles of a minimal three-generation model for neutrino
physics is relatively large, which opens the window for the discovery 
of leptonic CP violation.
This result exerts pressure on
flavor symmetry models (such as based on the permutation groups $A_4$ and $S_4$), as they are generically constructed
in a way to give at leading order tri-bimaximal (or bi-maximal) mixing -- thus implying $\theta_{13}=0$ 
(see \cite{Altarelli:2010gt} for an extended review).
Corrections from the charged sectors or next-to-leading contributions to the neutrino mass matrix are generally 
invoked to correct such patterns and make the models compatible with the experimental data.

With the recent discovery of a bosonic resonance \cite{aad:2012gk,Chatrchyan:2012gu},
showing all the characteristics of the SM Higgs boson, a long search
might soon come to a successful end.
In contrast no conclusive hints for physics beyond the Standard Model (SM) have
been found \cite{:2012rz,Chatrchyan:2012te,atlas:2012qk,Chatrchyan:2012jx,:2012mfa,%
Aad:2012zx,atlas:2012vd,Chatrchyan:2012sv,CMS:2012mx,atlas:res}.
Therefore, 
large areas of the parameter space of the simplest SUSY models are excluded now and
the allowed mass spectra as well as the best fit values to the data are pushed to higher and higher 
values, see, \eg~\cite{Bechtle:2012zk,Buchmueller:2012hv} and \Refs\ therein.
This has lead to an increasing interest in the study of SUSY models which provide new features. For 
instance, models with light stops and higgsinos can more easily explain the recent findings
within the minimal-supersymmetric standard model (MSSM), see \eg\ \cite{Papucci:2011wy,Mahbubani:2012qq,Feng:2012jf,Baer:2012uy},
 while models with broken $R$-parity  
might be able to hide much better at the LHC 
\cite{Allanach:2012vj,Han:2012cu}. Moreover, the MSSM extension with a singlet Higgs boson \cite{Ellwanger:2011aa,Gunion:2012gc,Gunion:2012zd,Das:2012rr,Ross:2012nr,King:2012tr}
or an additional gauge group can easier explain a Higgs boson with a mass of 126 GeV 
than the MSSM 
\cite{Haber:1986gz,Drees:1987tp,Cvetic:1997ky,Zhang:2008jm,Hirsch:2011hg,OLeary:2011yq,Gogoladze:2012yf}.
Most of these models will be tested at the LHC after its energy upgrade to 14~TeV. 

Generic models to describe the smallness of neutrino mass, such as the 
standard type-I seesaw mechanism~\cite{Minkowski:1977sc,Yanagida:1979as,GellMann:1980vs,Mohapatra:1979ia},
imply new physics close to the GUT scale. 
In these seesaw models, one can integrate out the heavy mediators in order to obtain the famous
$d=5$ Weinberg operator~\cite{Weinberg:1979sa},
\be
\mathcal{L}_{\text{eff}}^{d=5} \propto \frac{1}{\Lambda_{\text{NP}}} (\overline{L^{c}} {\rm i} \tau^{2} H)\, (H {\rm i} \tau^2 L)
\label{equ:weinberg}
\ee
which leads, after Electroweak Symmetry Breaking (EWSB), to Majorana masses  $m_\nu \sim v^2/\Lambda_{\mathrm{NP}}$ for the neutrinos. Here $L$ and $H$ stand for the SM  lepton and Higgs doublets, respectively, $v$ for the Higgs vev, and $\Lambda_{\mathrm{NP}}$ for the new physics scale.  For order one couplings, $\Lambda_{\mathrm{NP}}$ points to the GUT scale, which means that these models are not accessible by experiments. In addition, they contain more parameters than there are observables in the neutrino sector. 

It is therefore potentially attractive to test neutrino mass models with new particles at the TeV scale, since  current and future LHC upgrades can directly discover or constrain these models. In such models,  the smallness of the neutrino mass requires (at least) an additional suppression mechanism:
 Examples are radiative mass generation, where loop
suppression factors enter, see Refs.~\cite{Ma:1998dn,Babu:2001ex,deGouvea:2007xp,Bonnet:2012kz,Farzan:2012ev,Angel:2012ug} for systematic approaches, or models with a small lepton number
violating contribution, such as the inverse seesaw mechanism~\cite{Mohapatra:1986bd}
or SUSY with R-parity violation  where both aspects enter, see for example
\cite{Hempfling:1995wj,Nilles:1996ij,Borzumati:1996hd,Nardi:1996iy,Chun:1998gp,Hirsch:2000ef,%
Diaz:2003as,Dedes:2006ni,Allanach:2007qc,Ghosh:2008yh}. 

In this paper, we focus on another possibility, which has been recently drawing some attention:
If the operator in \equ{weinberg} is forbidden, such as by a discrete symmetry, a higher dimensional operator may dominate~\cite{Babu:1999me,Babu:2001ex,Chen:2006hn,Gogoladze:2008wz,Giudice:2008uua,Babu:2009aq,Gu:2009hu,Bonnet:2009ej,Picek:2009is,Liao:2010rx,Liao:2010ku,Liao:2010ny,Kanemura:2010bq}.
Such models can also be combined with supersymmetry, see for example
\cite{Gogoladze:2008wz,Krauss:2011ur}. The simplest possibilities of operators in the MSSM, leading to neutrino mass, might be
\begin{align} 
\mathcal{L}^{d=2n+5}_{\text{eff}} =  \frac{1}{\Lambda_{\mathrm{NP}}^{d-4}}
	(LLH_{u} H_{u}) (H_{d} H_{u})^{n} \, , \quad n=1,2,3, \hdots \, ,
\label{equ:hh}
\end{align}
whereas in the NMSSM,
\begin{align}
 \mathcal{L}^{d=n+5}_{\text{eff}} =  \frac{1}{\Lambda_{\mathrm{NP}}^{d-4}} (LLH_uH_u) (S)^{n} \, , \quad n=1,2,3, \hdots 
\label{equ:s}
\end{align}
or mixed combinations with \equ{hh} are possible, see \Ref~\cite{Krauss:2011ur} for a detailed discussion. In all these cases, neutrino mass is suppressed by a higher power of $\Lambda_{\text{NP}}$, which means that $\Lambda_{\text{NP}}$ may be potentially the TeV scale. 

The simplest possibility which works both in the MSSM and the NMSSM is the $d=7$ operator
\begin{align} 
\mathcal{L}^{d=7}_{\text{eff}} =  \frac{1}{\Lambda_{\mathrm{NP}}^{3}}
	(LLH_{u} H_{u}) (H_{d} H_{u}) \, .
\label{equ:hh7}
\end{align}
This operator is interesting for two reasons: First of all, it already implies a substantial suppression of neutrino mass, more than the $d=6$ operator in \equ{s} (for $n=1$). And second, in the NMSSM framework
the concept of neutrino mass generation can be extended to higher ($d>7$) dimensions, \cf, \equ{hh}, whereas \equ{s} as leading contribution to neutrino mass is in fact limited to $n \le 2$, and operators with mixed combinations of $S$ and $(H_{d} H_{u})$ cannot exceed $d=7$ as well.\footnote{The reason for that in the NMSSM is that the $\hat S \hat H_u \hat H_d$ and $\hat S^3$ terms have to be allowed by the discrete symmetry, which implies that certain lower dimensional operators leading to neutrino mass cannot be forbidden~\cite{Krauss:2011ur}.} Therefore, we focus on \equ{hh7} in this study.

In \Ref~\cite{Krauss:2011ur}, we have studied a particular example for the generation
of this operator which leads potentially to interesting
signals at the LHC. In this paper we go a step further and investigate
how potential GUT completions of such models look like and in which
$SU(5)$ representation the corresponding mediators would reside in. This is
motivated by the fact that some of the newly postulated particles
are charged under the SM gauge group and, thus, would destroy the successful
MSSM prediction for gauge coupling unification 
\cite{Dimopoulos:1981yj,Ibanez:1981yh,Marciano:1981un,Einhorn:1981sx,%
Amaldi:1991cn,Langacker:1991an,Ellis:1990wk}.
This is similar to the case of the seesaw type II and type III mechanisms
in which the corresponding $SU(2)$ triplets
have to be included in complete $SU(5)$ multiplets leading to interesting
effects for collider physics and the abundance of the relic dark matter
density
\cite{Rossi:2002zb,Buckley:2006nv,Hirsch:2008gh,Esteves:2009vg,%
Borzumati:2009hu,Esteves:2010ff,Biggio:2010me,Biggio:2012wx,%
Hirsch:2012yv,FileviezPerez:2012gg}. 
The requirement to obtain gauge coupling unification implies that not only
the mediators for the higher dimensional neutrino mass operators have masses
at the TeV scale, but also the members of the corresponding $SU(5)$ multiplets.
However, renormalization group effects can lead to a sizable mass splitting
in particular for those particles charged under $SU(3)_C$. 

The plan of this paper is as follows: we start with the model presented
in \Ref~\cite{Krauss:2011ur} and discuss its GUT completion considering
both, the MSSM Higgs sector as well as the NMSSM Higgs sector. 
As we will see, we have to embed heavy $SU(2)$ leptons, required by the model,
into $SU(5)$ five-plet representations, which implies the prediction of
new heavy d-quarks at the TeV-scale.
In order to avoid the $\mu$-problem  of the MSSM, we discuss the NMSSM
implementation in section~\ref{sec:NMSSM} and its challenges, as well
as we propose an origin of deviations from tri-bimaximal mixings in
this approach.
In section \ref{sec:dprime}, we  review the LHC and cosmological 
constraints on the heavy d-quarks, and propose fast enough decay modes
compatible with the model. 
In section \ref{sec:decompositions} we systematically
discuss the GUT completions of the possibilities for neutrino mass
from the higher dimensional operator suggested in
\cite{Krauss:2011ur}. Here we show that several of them would imply
non-perturbative gauge couplings already significantly below the GUT scale.
We constrain ourselves to the $SU(5)$ case as it is also a
subgroup of larger groups such as $SO(10)$ or $E_6$.
Finally in section~\ref{sec:conclusions} we summarize and draw our 
conclusions.

\section{A low energy model for a realization of a $d=7$ neutrino
mass operator and  its GUT completion}
\label{sec:our_model}

In \Ref~\cite{Ma:1998dn}, it has been demonstrated that at tree-level,
there are exactly three decompositions of the Weinberg operator,
which are known as type I, type II, and type III seesaw mechanisms, respectively.
Similarly, the $d=7$ operator in \equ{hh7} can be decomposed
systematically at tree-level, as discussed in \Ref~\cite{Bonnet:2009ej} for the
Two Higgs Doublet Model and in \Ref~\cite{Krauss:2011ur} for SUSY.
As a side remark, note that supersymmetry constrains the
possible decompositions of these operators in contrast to the Weinberg operator,
where all three possibilities have also a supersymmetric realization.

\begin{figure}[t]
\begin{center}
\includegraphics[width=.4\linewidth]{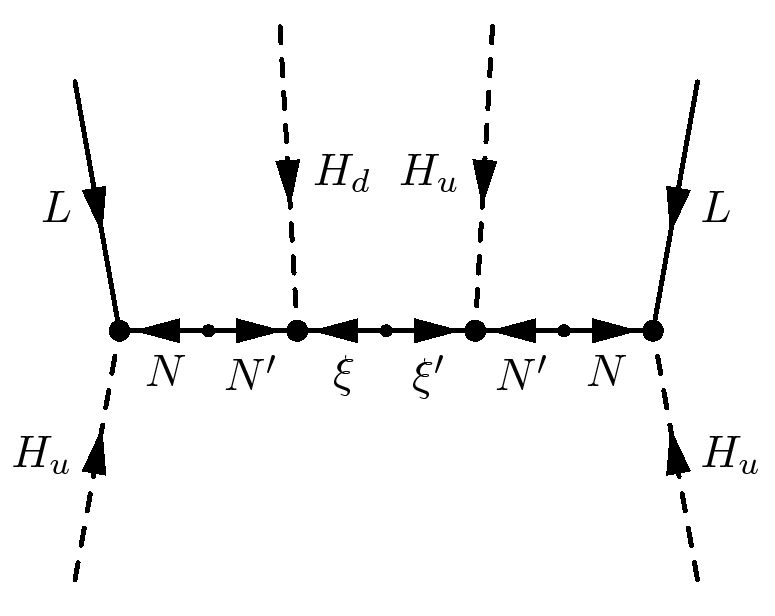}
\end{center}
\caption{\label{fig:example} Example for neutrino mass chosen in section~\ref{sec:our_model}. Here $N$ and $N'$ are fermion singlets, and $\xi$ and $\xi'$ are fermion doublets.}
\end{figure}

Here we will first recall a particular  decomposition of a $d=7$ operator, which has
been already presented in \Ref~\cite{Krauss:2011ur}, and discuss
its GUT completion to exemplify the challenges one encounters. The tree-level diagram
for the leading contribution to neutrino mass is shown in \figu{example}.
The corresponding superpotential is, at the electroweak scale, given by
\begin{align} \label{eq:W_our_model}
      W = \ & W_\text{\tiny{NMSSM}}
      + Y_N \hat N \hat L \cdot \hat H_u
      - \kappa_1 \hat N' \hat \xi \cdot \hat H_d
      + \kappa_2 \hat N' \hat \xi' \cdot \hat H_u
      + m_N \hat N \hat N' 
      + m_\xi \hat \xi \cdot \hat \xi'\,,
\end{align}
where the mediators $N$ and $N'$ are SM singlets and $\xi$ and
$\xi'$ are $SU(2)_L$ doublets carrying hypercharge $1/2$ and $-1/2$, respectively.
After electroweak symmetry breaking the mass matrix for the neutral weakly
interacting fermions reads schematically
in the basis $f^0 = (\nu, N, N',\xi^0, {\xi'}^0)$ as
\begin{align}
\label{equ:neutralfermions}
 &M^0_f =
\left(\begin{array}{ccccc}
	0			& Y _N v_u	& 0			& 0			& 0			\\
	Y_N v_u		& 0			& m_N		& 0			& 0			\\
	0			& m_N		& 0			& \kappa_1 v_d	& \kappa_2 v_u	\\
	0			& 0			& \kappa_1 v_d	& 0			& m_\xi		\\
	0			& 0			& \kappa_2 v_u	& m_\xi		& 0			
\end{array}\right)\,,
\end{align}
where we have neglected all flavor indices.
By integrating out the heavy fields one obtains an effective mass matrix for the 
three SM neutrinos
\begin{eqnarray}
m_\nu = v_u^3 v_d Y_N^2 \frac{\kappa_1 \kappa_2}{m_\xi m_N^2},\label{numass}
\end{eqnarray}
where the couplings carry a flavor index. As has been demonstrated in \Ref~\cite{Krauss:2011ur}, one successfully explains neutrino masses and mixings within this
model by an appropriate choice of fermion generations and coupling matrices. In addition, one can easily see from \equ{neutralfermions} that the neutral fermion mass matrix reduces to an inverse seesaw form if the $\xi^0$ and ${\xi'}^0$ are integrated out, where the $\hat \mu$-term (the 3-3-element in the inverse seesaw mass matrix) is suppressed by $m_\xi$.

Choosing the masses of the mediator fields at the TeV scale implies that
the additional couplings are $\mathcal O (10^{-3})$, which is comparable to the SM
Yukawa couplings. Accordingly, this model can be tested at the LHC. The
SM singlet fields $N$ and $N'$ are only produced in small amounts due
to the smallness of the Yukawa couplings.  The SU(2) doublets $\xi$
and $\xi'$, however, can be produced in Drell-Yan processes, similarly
to charginos and neutralinos, with a cross-section of up to ${\cal O}(10^2$ fb).
 These particles will then decay into vector bosons and
leptons which is induced by the mixing between the neutral fermions as described
by the mass matrix in \equ{neutralfermions}. The smallness of the
mixing between the heavy and the light neutrinos implies that one has to expect
sizable decay length of up to several millimeters. Under favorable
conditions one can even establish connections between neutrino
physics and lepton number violating LHC signals arising
from the production of two same sign charged leptons and several
$W$-bosons.

The presence of the $\xi$ and $\xi'$ 
supermultiplets modifies the running of the gauge couplings such that
unification would not occur anymore. Therefore, one has to embed them
into complete representations of a GUT gauge group.\footnote{In certain models it is possible to conserve gauge coupling unification without having complete multiplets of the GUT gauge group~\cite{Calibbi:2009cp}. The introduction of additional fields, however, is also required in these cases. } In the following we will focus on gauge groups that contain SU(5) as a subgroup. As the doublet mediators carry the same quantum
numbers as the MSSM Higgs bosons the natural choices are $5_{\xi}$ and
$\bar{5}_{\xi'}$. The minimal field content is given by
\begin{equation}
\bar 5_M=\left( \begin{array}{c} (d_R)^c \\ L \end{array} \right) \, , \quad
\bar 5_{\xi'}=\left( \begin{array}{c} d'^c \\ \xi' \end{array} \right) \, , \quad
5_{\xi}=\left( \begin{array}{c} d'' \\ \xi \end{array} \right) \, , \quad
H_5 = \left( \begin{array}{c} H_{\text{col}} \\ H_u \end{array} \right)  \, , \quad
H_{\bar 5} = \left( \begin{array}{c} H'_{\text{col}} \\ H_d \end{array} \right) \, , \quad
\end{equation}
and the fermionic singlets $N$ and $N'$.
This implies that at the electroweak scale one has
not only the new leptonic fields $\xi'$ and $\xi$, but also additional heavy quarks
$d'^c$ and $d'{}'$ stemming from $\bar{5}_{\xi'}$ and $5_{\xi}$,
respectively. We will discuss their phenomenology in section  \ref{sec:dprime}.
For completeness we recall that the fields $N$ and $N'$ are (fermionic)
gauge singlets, similar to the NMSSM singlet scalar field $S$.
Note that in SUSY SU(5) models one usually needs additional Higgs 42-plets in order to describe the quark masses correctly. These 42-plets, however, are not relevant for the following discussion.

The most general $SU(5)$  invariant superpotential
containing the MSSM fields, $\bar{5}_{\xi'}$, $5_{\xi}$, $N$ and $N'$
reads as
\begin{eqnarray} \label{eq_W_MSSM}
W&=& y_1\,N \, 5_\xi\,H_{\overline 5}+y_2\,N \, \overline{5}_{\xi^\prime}\,H_5+y_3\,N \,\overline{5}_M\,H_5
+\nonumber \\
&&y_1^\prime,N^\prime \, 5_\xi\,H_{\overline 5}+y_2^\prime\,N^\prime \, \overline{5}_{\xi^\prime}\,H_5+y_3^\prime\,N^\prime \,\overline{5}_M\,H_5 + \nonumber \\
&&m_{\xi^\prime}\, \overline{5}_M\, 5_\xi +
m_{\xi}\, \overline{5}_{\xi^\prime}\, 5_\xi + m_N N' N + m_{NN} N N + m_{N'N'} N' N' + \nonumber \\
&& y_d\,\overline{5}_M\,10\,H_{\overline 5} +  y_d^\prime\,
\overline{5}_{\xi^\prime}\,10\,H_{\overline 5} + y_u\,10\,10\,H_5
 - \mu H_{\bar 5} H_5\,.
\end{eqnarray}
If both $y_3$ and $y_3'$ were present, then one would generate the
Weinberg operator which in general would be the dominant source
for neutrino masses. However, if there were a discrete
flavor symmetry operating at the GUT scale, then this operator
could be forbidden, implying that the $d=7$ operator in \equ{hh7} is the dominant
contribution to neutrino mass. As an explicit example we have taken 
 a
$\mathds{Z}_3$ symmetry with the charge assignments given
in \Tab~\ref{tab:charges}.
\begin{table}[t]
\caption{\label{tab:charges}
 Possible $\mathds{Z}_3$ assignments to forbid the
Weinberg operator.}
\begin{center}
\begin{tabular}{c||cccccccc}
Multiplet		& $\bar 5_M$	& $H_5$	& $H_{\bar 5}$	& $N$	& $N'$	& $5_\xi$	& $\bar 5_{\xi'}$	& $10$  \\ \hline
$\mathds{Z}_3$ charge	& 1		& 1	& 1		& 1	& 2	& 0		& 0		 	& 1
\end{tabular}
\end{center}
\end{table}
The superpotential hence is then reduced to
\begin{eqnarray} \label{eq:SU5W_reduced}
W&=& y_3\,N \,\bar{5}_M\,H_5 +
y_1^\prime,N^\prime \, 5_\xi\,H_{\bar 5}+y_2^\prime\,N^\prime \, \bar{5}_{\xi^\prime}\,H_5+ \nonumber\\
&&m_{\xi}\, \bar{5}_{\xi^\prime}\, 5_\xi + m_N N' N \nonumber\\
&& y_d\,\bar{5}_M\,10\,H_{\bar 5} +
y_u\,10\,10\,H_5 - \mu H_{\bar 5} H_5 \,.
\end{eqnarray}
The superpotential in \eq~(\ref{eq:W_our_model}) is obtained identifying $y_3 \widehat= Y_N$, $y_1' \widehat= \kappa_1$
and $y_2' \widehat= \kappa_2$. Note that the term $\mu H_{\bar 5} H_5$ breaks the discrete symmetry explicitly (see discussion in \Refs~\cite{Bonnet:2009ej,Krauss:2011ur}). Furthermore, a somewhat problematic feature
of this approach is, that the $\mu$-problem of the MSSM is enlarged
as not only $\mu$, but also $m_{\xi}$ and $m_N$ have to be few hundred GeV
up to at most a few TeV.

Usually, one argues that the $\mu$ problem can be evaded if one enlarges
the MSSM Higgs sector by a singlet $S$. This leads to the NMSSM,
where $\mu = \lambda \langle v_S \rangle$ and $\lambda$ is the
coupling between the MSSM Higgs doublets and  $S$. 
Indeed, one could in principle also obtain $m_{\xi}$ and $m_N$ in this
way in the desired order of magnitude. However, as we will discuss
in the next section, this contradicts the requirement that
the Weinberg operator should be forbidden.

\section{Implementation in the NMSSM, and the origin of a non-zero $\theta_{13}$?}
\label{sec:NMSSM}

In the NMSSM, one introduces an extra gauge singlet $S$ which couples to the
Higgs doublets such that $\mu = \lambda \langle S \rangle$. Similarly one
can generate in this way also the mass terms for additional particles in the
model discussed above. For this, the generalization of the superpotential in 
\eq~(\ref{eq_W_MSSM}) reads as
\begin{eqnarray}
W&=& y_1\,N \, 5_\xi\,H_{\overline 5}+y_2\,N \, \overline{5}_{\xi^\prime}\,H_5+y_3\,N \,\overline{5}_M\,H_5
+\nonumber \\
&&y_1^\prime,N^\prime \, 5_\xi\,H_{\overline 5}+y_2^\prime\,N^\prime \, \overline{5}_{\xi^\prime}\,H_5+y_3^\prime\,N^\prime \,\overline{5}_M\,H_5 + \nonumber \\
&&\lambda_{\xi^\prime}\, S\, \overline{5}_M\, 5_\xi +
\lambda_{\xi}\, S\, \overline{5}_{\xi^\prime}\, 5_\xi + \lambda_N S\, N' N + \lambda_{NN} S\, N N + \lambda_{N'N'} S\, N' N' +  \nonumber \\
&& y_d\,\overline{5}_M\,10\,H_{\overline 5} +  y_d^\prime\,
\overline{5}_{\xi^\prime}\,10\,H_{\overline 5} + y_u\,10\,10\,H_5 \,.
\end{eqnarray}
After electroweak symmetry breaking, the heavy mass scale is automatically set by
$\VEV{S}$, which is of order of TeV. This has immediate consequences for the
effective neutrino mass operators, which have now the schematic
structure
\begin{align*}
 &\frac{1}{\langle{S}\rangle} LLH_uH_u, \qquad
 &&\frac{1}{\langle{S}\rangle^3} (LLH_uH_d)(H_uH_d)\,.
\end{align*}
As a consequence $\langle S \rangle$ breaks any discrete symmetry
under which it is charged. This prevents one from choosing a simple
discrete symmetry group to avoid the $d=5$ contribution to neutrino
mass. Especially the term $\lambda_{\xi^\prime}\, S\, \overline{5}_M\,
5_\xi$, which leads to a mixing between the light and heavy mediators
is problematic in this context, as it induces the Weinberg operator.

We will prove now that one cannot forbid
this term by a discrete symmetry that fulfills the required
conditions. For this we
start with three (yet) unconstrained charges as parameters
\begin{align}
 q_S \equiv s \,,\quad q_{H_5} \equiv 2 h' \,,\quad q_N = n \,.
\end{align}
From the absolutely necessary terms in the superpotential  we derive
\begin{subequations}
\begin{align}
 &(S H_5 H_{\bar 5}) 	&&\Rightarrow &q_{H_{\bar 5}} &= - s - 2 h' \\
 &(S N N') 		&&\Rightarrow &q_{N'} &= - s - n \\
 &(N \bar 5_M H_5)	&&\Rightarrow &q_M &= -2h' - n \\
 &(\bar 5_M 10 \bar H_{\bar 5})	&&\Rightarrow &q_{10} &= 4 h' + n + s\,.
\end{align}
\end{subequations}
From the term $(10 10 H_5)$ we obtain
\begin{align}
 n = -s - 5h'\,.
\end{align}
which leads to the following set of equations
\begin{equation}
 q_{H_{\bar 5}} = - s - 2 h' \,\,,\,\,
 q_{N'} = 5 h' \,\,,\,\,
 q_M = 3 h' + s \,\,,\,\,
 q_{10} = -h'\,.
\end{equation}
As a consequence we derive for the charges of the doublets
\begin{align}
&(N' H_{\bar 5} 5_\xi) 	&&\Rightarrow & q_\xi &= -3 h' + s \\
&(N' \bar 5_{\xi'} H_5)	&&\Rightarrow & q_{\xi'} &= -7 h' \\
\end{align}
leading to
\begin{align}
 q(S \bar 5_M 5_\xi) = 3s \,,
\end{align}
but we know that $3s = 0$ since we need a term $S^3$ in the NMSSM.
This implies that one cannot forbid this unwanted term. Note, that 
this holds for every Abelian discrete symmetry group. As a consequence one can show along
the same lines that every higher dimension operator for neutrino masses
\begin{equation}
\mathcal{O}^{d} = \frac{1}{\VEV{S}^{1+2k+l}}(LLH_uH_u)(H_uH_d)^kS^l
\end{equation}
has to have the same charge as the Weinberg operator.
We have also checked that similar problems appear if the 
Abelian symmetry is chosen as a product of two different cyclic groups 
$Z_N \otimes Z_{N^\prime}$. 

The next possibility is the 
addition of a second singlet $S'$ but this does not improve the situation either as one
has to have interactions such that also $S'$ obtains a VEV. However, 
from the term $SSS'$
we obtain immediately $q_{S'} = q_S$.
As a consequence we find that the symmetry must  be broken, which
implies that also the Weinberg operator can be induced.
 However, if this breaking is small enough, then  the resulting Weinberg operator
contributes only sub-dominantly to the neutrino masses and mixing angles.

\begin{table}
\begin{center}
\begin{tabular}{c||cccccccccc}
Multiplet		& $\bar 5_M$	& $H_5$	& $H_{\bar 5}$	& $N$	& $N'$	& $5_\xi$	& $\bar 5_{\xi'}$	& $10$	& $S$	& $S'$\\ \hline
$\mathds{Z}_3$ charge	& 1		& 1	& 1		& 1	& 2	& 0		& 0		 	& 1	& 1	& 0
\end{tabular}
\end{center}
\caption{\label{tab:model2}
 Charges for the fields of the model defined in \eq~(\ref{eq:borkenmodel}).}
\end{table}
We consider the above model extended by a $S'$ to exemplify the main points. The charges of
the particles for the case of an unbroken symmetry are given in \Tab~\ref{tab:model2} and 
the superpotential\footnote{Note that this charge assignment would allow for a 
quadratic mass term for $S'$. However, we only allow for trilinear terms, which can 
be easily achieved, \eg, by adding an additional $\mathds{Z}_3$ under which all 
fields have the same charge.} at the $SU(5)$ level is given by 
\begin{align}
 \begin{split}
  W=& \quad y_3\,N \,\overline{5}_M\,H_5
+ y_1^\prime,N^\prime \, 5_\xi\,H_{\overline 5}+y_2^\prime\,N^\prime \, \overline{5}_{\xi^\prime}\,H_5 
 + \lambda_{\xi}\, S^\prime \, \overline{5}_{\xi^\prime}\, 5_\xi + \lambda_N S^\prime \, N' N \\
& + y_d\,\overline{5}_M\,10\,H_{\overline 5} + y_u\,10\,10\,H_5
 + \lambda_S S H_{\overline{5}} H_5 + \kappa S^3  + \kappa'  {S^\prime}^3 \\
& + \lambda_S^\prime S^\prime H_{\overline{5}} H_5 +  y_3^\prime\,N^\prime \,\overline{5}_M\,H_5
+ y_d^\prime \,\overline{5}_\xi\,10\,H_{\overline 5} + \dotsb
\,,
 \end{split}
 \label{eq:borkenmodel}
\end{align}
where the interactions in the first two lines respect the discrete symmetry and the ones of the last
line break it. The $\lambda_S^\prime S^\prime H_{\overline{5}} H_5$ term is a possibility  to obtain a VEV 
for $S'$ such that the mediators
for the neutrino mass operators have their masses proportional to $\VEV{S'}$. The second term of this
line yields the Weinberg operator after integrating out the heavy fields:
\begin{eqnarray}
 m_\nu^{d=5} = \frac{y_3 y_{3}' v_u^2}{\lambda_N\VEV{S'}}\,. \label{eq:mnu_corr}
\end{eqnarray}
We obtain from the requirement
\begin{align}
m_\nu^{d=5}  \ll m_\nu^{d=7} = \frac{y_1' y_2' y_3^2 v_u^3 v_d}{\lambda_N^2 \lambda_\xi \VEV{S'}^3}\,, \label{eq:mnu_7}
\end{align}
that $y_3^\prime < 10^{-8}$ assuming (symmetry conserving) couplings of the order $10^{-2}$ and $\VEV{S'}\sim$~TeV.
 Hence we require all symmetry breaking couplings to be of order $10^{-8}$ or smaller. We have
 checked that $\lambda_i\VEV{S'}$ is of $O($TeV$)$ in spite of $\lambda_S^\prime$ being so small.

It is interesting to observe that, for the symmetry breaking coupling $y_3'$ close to this upper limit, 
the neutrino mass matrix in \eq~(\ref{eq:mnu_7}) could receive a new (and potentially large) contribution from the $d=5$ operator in \eq~(\ref{eq:mnu_corr}). Suppose that the $d=7$ operator has a tri-bimaximal flavor structure, where (for the sake of simplicity) the charged lepton mass matrix is assumed to be diagonal. Then the deviation from tri-bimaximal mixing can originate in the symmetry breaking terms in \eq~(\ref{eq:borkenmodel}). 
We illustrate  this effect with an exemplary flavor structure, not discussing the details of the underlying flavor symmetry. For reason of simplicity,  we use a minimal scenario where the mass of the lightest neutrino is zero. 
For the $d=7$ contribution, we assume a flavor structure where we have three generations of the singlet fields $N$ and $N'$ and the following structure of the coupling constants
\footnote{Here we assume a similar flavor structure as in \Ref~\cite{Krauss:2011ur} 
but three instead of two generations of $N$ and $N'$.}
\begin{align}
y_1' = \widetilde y_1'
\begin{pmatrix}
 0 \\ 1 \\ \rho
\end{pmatrix} \,,\quad
y_2' = \widetilde y_2'
\begin{pmatrix}
 0 \\-1 \\ \rho
\end{pmatrix} \,,\quad
 y_3 = \widetilde y_3
\begin{pmatrix}
\sqrt{2/3} & 1 / \sqrt{3} & 0 \\
-1/\sqrt{6} & 1/\sqrt{3} & -1/\sqrt{2} \\
-1/\sqrt{6} & 1/\sqrt{3} & 1/\sqrt{2} 
\end{pmatrix}\,,
\end{align}
where $\widetilde y_1$, $\widetilde y_2$ and $\widetilde y_3$ are numerical (scalar) parameters, $\rho = \sqrt{m_3/m_2}$ and the mass matrices for $N$ and $N'$ are diagonal. This choice of parameters will generate a tri-bimaximal mass matrix for the neutrinos at $d=7$
\begin{align}\begin{split}
m_\nu^{d=7} &= \frac{v_u^3 v_d}{\VEV{S}^3} \cdot y_3 \left[y_1'(y_2')^\T + y_2'(y_1')^\T \right]y_3^\T\\
&= U_\text{TB}\cdot\mathrm{diag}(0,m_2,-m_3)\cdot U_\text{TB}^\T\,,
\end{split}
\end{align}
where $U_\text{TB}$ is the tri-bimaximal mixing matrix.
We can write the total neutrino mixing matrix including small linear deviations from tri-bimaximal mixing by using the parametrization from \Ref~\cite{King:2007pr} (note the different sign convention)
\begin{align}
U_\text{PMNS} =
\begin{pmatrix}
 \sqrt{\frac{2}{3}}(1-\frac{1}{2}s) & \frac{1}{\sqrt{3}}(1+s) & \frac{1}{\sqrt{2}} r \\
 -\frac{1}{\sqrt{6}}(1+s-a-r) & \frac{1}{\sqrt{3}}(1-\frac{1}{2}s -a +r) & -\frac{1}{2}(1+a) \\
 -\frac{1}{\sqrt{6}}(1+s+a+r) & \frac{1}{\sqrt{3}}(1-\frac{1}{2}s +a -r) & \frac{1}{2}(1-a)  
\end{pmatrix}\, ,
\end{align}
where $\sin \theta_{13} = r/\sqrt{2}$, $\sin \theta_{12} = 1/\sqrt{3} (1+s)$, and $\sin \theta_{23}=1/\sqrt{2} (1+a)$.
Since we want to generate the corrections to the neutrino mass matrix by the $d=5$ contribution, we can write:
\begin{align}
 m_\nu &= m_\nu^{d=7} + m_\nu^{d=5} \, , \quad \text{or} \\
U_\text{PMNS}\cdot\mathrm{diag}(0,m_2,-m_3)\cdot U_\text{PMNS}^\T &=  U_\text{TB}\cdot\mathrm{diag}(0,m_2,-m_3)\cdot U_\text{TB}^\T + \frac{v_u^2}{\VEV{S}} (y_3 (y_3')^\T + y_3' (y_3)^\T)
\end{align}
Solving this equation for the symmetry breaking coupling $y_3'$, we obtain
\begin{align}
\begin{split}
 y_3'\simeq \widetilde y_3'
 \begin{pmatrix}
 \frac{s}{\sqrt{6}} & \frac{s}{\sqrt{3}} & \frac{(2a-r) -3r\,\rho^2}{3 \sqrt{2}}  \\
  \frac{s+(a+r)\rho^2}{\sqrt{6}} & \frac{-(2 a+r-s)-(2a-r)\rho^2}{2 \sqrt{3}} & \frac{(2a-r)+3a\,\rho^2}{3 \sqrt{2}} \\
 \frac{s\, - (a+r)\rho^2}{\sqrt{6}} & \frac{(2 a-r-s) + (2a-r) \rho^2}{2 \sqrt{3}} & \frac{(2a-r)+3a\,\rho^2}{3 \sqrt{2}}
\end{pmatrix}
\end{split}
\end{align}
to first order in $r$, $s$, and $a$. If we want to consider only corrections to $\theta_{13}$ (\ie, $a=0$, $s=0$), we have, 
\begin{align}
\begin{split}
 y_3'\simeq \widetilde y_3' \, r \, \left[
 \begin{pmatrix}
0 & 0 & - \frac{1}{ \sqrt{2}} \\
   \frac{1}{\sqrt{6}} &  \frac{1}{2 \sqrt{3}} & 0 \\
 - \frac{1}{\sqrt{6}} &- \frac{1}{2 \sqrt{3}} & 0
\end{pmatrix} \rho^2
- \begin{pmatrix}
   0 & 0 & \frac{1}{3\sqrt{2}} \\
   0 & \frac{1}{2\sqrt{3}} & \frac{1}{3\sqrt{2}} \\
   0 & \frac{1}{2\sqrt{3}} & \frac{1}{3\sqrt{2}}
  \end{pmatrix}
\right]
\end{split}
\end{align}
with $\mathcal{O}(\widetilde y_3') = 10^{-8}$. This will give a correction to the tri-bimaximal mass matrix from the $d=7$ operator that leads to the observed neutrino mixing with $\theta_{13} > 0$. In a specific model, such a structure may originate from a flavor symmetry.

\section{Phenomenology of the additional $d$-quarks}
\label{sec:dprime}

The promotion of the additional leptonic SU(2) doublets $\xi$ and
$\xi'$ to 5-plets $5_\xi$ and $\bar{5}_{\xi'}$ of SU(5) implies the
existence of additional $d$-quarks at the TeV scale as one wants to
maintain gauge coupling unification. We will denote the corresponding
mass eigenstates by $D'$, which is composed of  $d^c{}'$ and
 $d'{}'$, and $L'$ which is composed of $\xi$ and $\xi'$.
At the GUT-scale $L'$ and $D'$ have a common mass $m_{\xi,GUT}$.
Renormalization group effects will lead to shift of the $L'$ mass 
$m_\xi$ and the $D'$ mass $m_{D'}$.  
At one-loop order the corresponding RGEs read as
\begin{equation}
\frac{d}{dt} m_k = \sum_i \frac{c^k_i \alpha_i(t)}{4 \pi} m_k \hspace*{2cm}
(k=\xi,D')
\end{equation}
where the index $i$ runs over the different gauge groups, $t=\ln (Q^2/M^2_{GUT})$, 
$c^{D'}_i=(-4/15,0,-16/3)$  and $c^\xi_i=(-3/5,-3,0)$ for
$i=U(1)_Y, SU(2)_L,SU(3)_C$.  Here we have assumed that one can
neglect the Yukawa interactions compared to the gauge interactions. 
These equations can easily be solved and one obtains
\begin{equation}
m_k(t) = \prod_i \left( \frac{\alpha_i(t)}{\alpha_{GUT}}\right)^\frac{c^k_i}{b_i} 
m_{\xi,GUT}
\end{equation}
with $b_i=(38/5,2,-2)$.
Clearly the precise ratio of these two masses at the electroweak scale 
depends on the particle content between the
electroweak scale and the GUT scale as this changes the $b_i$. In our
model we obtain $m_{D'}/m_{\xi} \simeq 5$ at $Q=1$~TeV. 
Provided that these leptons can be observed at LHC up to mass of about
800 GeV \cite{delAguila:2008iz,Krauss:2011ur} the $D'$ can have
a mass of up to 4 TeV whereas LHC might observe them up to a mass of about
3 TeV  see \eg~\cite{delAguila:2008iz} and \Refs\
therein.

Even though these quarks are heavier than the additional leptons,
they will be stable in the model discussed so far as these particles
are protected by the  same symmetry that forbids the $d=5$ operator. Stable heavy
d-quarks, however, cause conflicts with cosmological constraints.
Their presence during Big Bang Nucleosynthesis (BBN) would alter the
observed abundances of the light elements in the universe (see, \eg,
\Ref~\cite{Iocco:2008va} for a review). Further bounds come, \eg,
from direct heavy element searches in water~\cite{Beringer:1900zz}.
Although these additional quarks can annihilate via gluons
into SM quarks and gluons, it turns out this mechanism is not sufficient
to lower the yield below the experimental requirements
 \cite{Nardi:1990ku,Berger:2008ti} if their masses are below 2.5 TeV, \eg\ the range
interesting for LHC. 
To summarize, heavy stable d-quarks must have life-times much smaller than the age of the
 universe to avoid constraints from direct searches, and annihilation processes that 
are efficient enough to lower their abundance below the bounds from BBN. These bounds 
do not affect particles that decay before BBN ($\tau \ll 1\unit{s}$).
 
In order to let the $D'$ decay fast enough,  one therefore needs additional small terms that break the symmetry. For example the interactions of \eq~(\ref{eq:borkenmodel}), discussed in the context of the NMSSM,
 lead to the two-body decays 
\begin{align}
  D' \to H^-  u\,,\, H^0 d \ .
\label{eq:dprime_decay}
\end{align}
which result in life-times  as small as $10^{-10}\unit{s}$ to $10^{-13}\unit{s}$, depending on
$m_{D'}$, if the symmetry breaking couplings are of
order $10^{-8}$. The signal for such a small life-time at the LHC
is a displaced vertex. Measuring the 
corresponding decay length immediately gives the size of the corresponding coupling.
\section{Systematic review of the decompositions, and their SU(5) completions}
\label{sec:decompositions}

In the previous sections, we have discussed a specific realization
of a supersymmetric model
where the dominant contribution to neutrino masses is given
by a $d=7$ operator. The complete list of 
$d=7$ operators including
the various mediators is given in \cite{Krauss:2011ur}, where the previous model
is listed as \#17. For convenience
of the reader, we display here the possibilities in \Tab~\ref{tab_mediators}
where we also include the $SU(5)$ multiplets containing the mediators.

\begin{table}[p]
\begin{center}
{\small 
 \begin{tabular}{rccc}
\hline \hline
\# & Operator & Mediators & SU(5) multiplets 
\\
\hline \rule[0em]{0em}{1.2em}  
1
&
     $(H_{u} {\rm i} \tau^{2} \overline{L^{c}})
     (H_{u} {\rm i} \tau^{2} L)(H_{d} {\rm i}
     \tau^{2} H_{u})$
&
${\bf 1}^{R}_{0}$, ${\bf 1}^{L}_{0}$, ${\bf 1}^{s}_{0}$
&
 \med{1, 1, 1}
\\
2
&
     $(H_{u} {\rm i} \tau^{2} \vec{\tau} \overline{L^{c}})
     (H_{u} {\rm i} \tau^{2} L)
     (H_{d} {\rm i} \tau^{2} \vec{\tau} H_{u})$
&
${\bf 3}^{R}_{0}$, ${\bf 3}^{L}_{0}$
${\bf 1}^{R}_{0}$, ${\bf 1}^{L}_{0}$,
${\bf 3}^{s}_{0}$
&
 \med{24, 24, (1), (1), 24}
\\
3
&
     $(H_{u} {\rm i} \tau^{2} \vec{\tau} \overline{L^{c}})
     (H_{u} {\rm i} \tau^{2} \vec{\tau} L)
     (H_{d} {\rm i} \tau^{2} H_{u})$
&
${\bf 3}^{R}_{0}$, ${\bf 3}^{L}_{0}$, ${\bf 1}^{s}_{0}$
&
 \med{24, 24, 1}
\\
4
&
$ 
     (-{\rm i} \epsilon^{abc})
     (H_{u} {\rm i} \tau^{2} \tau^{a} \overline{L^{c}})
     (H_{u} {\rm i} \tau^{2} \tau^{b} L)
     (H_{d} {\rm i} \tau^{2} \tau^{c} H_{u})$
&
${\bf 3}^{R}_{0}$, ${\bf 3}^{L}_{0}$, ${\bf 3}^{s}_{0}$
&
 \med{24,24,24}
\\
  	5 
&
     $(\overline{L^{c}} {\rm i} \tau^{2} \vec{\tau} L) 
     (H_{d} {\rm i} \tau^{2} H_{u}) 
     (H_{u} {\rm i} \tau^{2} \vec{\tau}  H_{u})$
&
	${\mathbf 3}^{s}_{+1}$, ${\mathbf 3}^{s}_{+1}$, ${\mathbf 1}^{s}_{0}$ 
&
 \med{15,15,1}
\\
  \hline \rule[0em]{0em}{1.2em}  
 	6 
&
     $(-\ii\epsilon_{abc})(\overline{L^{c}} {\rm i} \tau^{2} \tau_a L) 
     (H_{d} {\rm i} \tau^{2} \tau_b H_{u}) 
     (H_{u} {\rm i} \tau^{2} \tau_c  H_{u})$
&
	${\mathbf 3}^{s}_{+1}$, ${\mathbf 3}^{s}_{+1}$, ${\mathbf 3}^{s}_{0}$ 
&
 \med{15,15,24}
\\
7
&
$(H_{u} {\rm i} \tau^{2} \overline{L^{c}})
(L {\rm i} \tau^{2} \vec{\tau} H_{d})
(H_{u} {\rm i} \tau^{2} \vec{\tau} H_{u})$
&
${\bf 1}^{R}_{0}$, ${\bf 1}^{L}_{0}$,
${\bf 3}^{R}_{-1}$, ${\bf 3}^{L}_{-1}$,
${\bf 3}^{s}_{+1}$
&
 \med{1,1,15,\overline{15},15}
\\
8
&
$(-{\rm i} \epsilon^{abc})
(H_{u} {\rm i} \tau^{2} \tau^{a} \overline{L^{c}})
(L {\rm i} \tau^{2} \tau^{b} H_{d})
(H_{u} {\rm i} \tau^{2} \tau^{c} H_{u})$
&
${\bf 3}^{R}_{0}$, ${\bf 3}^{L}_{0}$,
${\bf 3}^{R}_{-1}$, ${\bf 3}^{L}_{-1}$,
${\bf 3}^{s}_{+1}$
&
 \med{24,24,15,\overline{15},15}
\\
9
&
$(H_{u} {\rm i} \tau^{2} \overline{L^{c}})
 ({\rm i} \tau^{2} H_{u})
 (L)
 (H_{d} {\rm i} \tau^{2} H_{u})$
&
${\bf 1}^{R}_{0}$, ${\bf 1}^{L}_{0}$, 
${\bf 2}^{R}_{-1/2}$, ${\bf 2}^{L}_{-1/2}$,
${\bf 1}^{s}_{0}$
&
 \med{1,1,5,\overline{5},1}
\\
10
&
$(H_{u} {\rm i} \tau^{2} \vec{\tau} \overline{L^{c}})
 ({\rm i} \tau^{2} \vec{\tau} H_{u})
 (L)
 (H_{d} {\rm i} \tau^{2} H_{u})$
&
${\bf 3}^{R}_{0}$, ${\bf 3}^{L}_{0}$, 
${\bf 2}^{R}_{-1/2}$, ${\bf 2}^{L}_{-1/2}$,
${\bf 1}^{s}_{0}$
&
 \med{24,24,5,\overline{5},1}
\\
\hline \rule[0em]{0em}{1.2em}  
11
&
$(H_{u} {\rm i} \tau^{2} \overline{L^{c}})
 ({\rm i} \tau^{2} H_{u})
 (\vec{\tau} L)
 (H_{d} {\rm i} \tau^{2} \vec{\tau} H_{u})$
&
${\bf 1}^{R}_{0}$, ${\bf 1}^{L}_{0}$, 
${\bf 2}^{R}_{-1/2}$, ${\bf 2}^{L}_{-1/2}$,
${\bf 3}^{s}_{0}$
&
 \med{1,1,5,\bar{5},24}
\\
12
&
$(H_{u} {\rm i} \tau^{2} \tau^{a} \overline{L^{c}})
 ({\rm i} \tau^{2} \tau^{a} H_{u})
 (\tau^{b} L)
 (H_{d} {\rm i} \tau^{2} \tau^{b} H_{u})$
&
${\bf 3}^{R}_{0}$, ${\bf 3}^{L}_{0}$, 
${\bf 2}^{R}_{-1/2}$, ${\bf 2}^{L}_{-1/2}$,
${\bf 3}^{s}_{0}$
&
 \med{24,24,5,\bar{5},24}
\\
13
&
     $(H_{u} {\rm i} \tau^{2} \overline{L^{c}})(L) 
     ({\rm i} \tau^{2} H_{u})
     (H_{d} {\rm i} \tau^{2} H_{u})$
&
${\bf 1}^{R}_{0}$, ${\bf 1}^{L}_{0}$,
${\bf 2}^{s}_{+1/2}$, ${\bf 1}^{s}_{0}$
&
 \med{1,1,5,1}
\\
14
&
     $(H_{u} {\rm i} \tau^{2} \vec{\tau} \overline{L^{c}})
     (\vec{\tau} L) 
     ({\rm i} \tau^{2} H_{u})
     (H_{d} {\rm i} \tau^{2} H_{u})$
&
${\bf 3}^{R}_{0}$, ${\bf 3}^{L}_{0}$,
${\bf 2}^{s}_{+1/2}$, ${\bf 1}^{s}_{0}$
&
 \med{24,24,5,1}
\\
15
&
     $(H_{u} {\rm i} \tau^{2} \overline{L^{c}})(L) 
     ({\rm i} \tau^{2} \vec{\tau} H_{u})
     (H_{d} {\rm i} \tau^{2} \vec{\tau} H_{u})$
&
${\bf 1}^{R}_{0}$, ${\bf 1}^{L}_{0}$,
${\bf 2}^{s}_{+1/2}$, ${\bf 3}^{s}_{0}$
&
 \med{1,1,5,24}
\\
\hline \rule[0em]{0em}{1.2em}  
16
&
     $(H_{u} {\rm i} \tau^{2} \tau^{a} \overline{L^{c}})( \tau^{a} L) 
     ({\rm i} \tau^{2} \tau^{b} H_{u})
     (H_{d} {\rm i} \tau^{2} \tau^{b} H_{u})$
&
${\bf 3}^{R}_{0}$, ${\bf 3}^{L}_{0}$,
${\bf 2}^{s}_{+1/2}$, ${\bf 3}^{s}_{0}$
&
 \med{24,24,5,24}
\\
17
&
     $(H_{u} {\rm i} \tau^{2} \overline{L^{c}})
     (H_{d}) ({\rm i} \tau^{2} H_{u}) 
     (H_{u} {\rm i} \tau^{2} L)$
&
${\bf 1}^{R}_{0}$, ${\bf 1}^{L}_{0}$,
${\bf 2}^{R}_{-1/2}$, ${\bf 2}^{L}_{-1/2}$
&
 \med{1,1,5,\bar{5}}
\\
18
&
     $(H_{u} {\rm i} \tau^{2}\vec{\tau} \overline{L^{c}})
     (\vec{\tau} H_{d}) 
     ({\rm i} \tau^{2} H_{u}) 
     (H_{u} {\rm i} \tau^{2} L)$
&
${\bf 3}^{R}_{0}$, ${\bf 3}^{L}_{0}$,
${\bf 2}^{R}_{-1/2}$, ${\bf 2}^{L}_{-1/2}$,
${\bf 1}^{R}_{0}$, ${\bf 1}^{L}_{0}$
&
 \med{24,24,5,\bar{5},(1),(1)}
\\
19
&
     $(H_{u} {\rm i} \tau^{2} \overline{L^{c}})
     (H_{d}) ({\rm i} \tau^{2} \vec{\tau} H_{u}) 
     (H_{u} {\rm i} \tau^{2} \vec{\tau} L)$
&
${\bf 1}^{R}_{0}$, ${\bf 1}^{L}_{0}$,
${\bf 2}^{R}_{-1/2}$, ${\bf 2}^{L}_{-1/2}$,
${\bf 3}^{R}_{0}$, ${\bf 3}^{L}_{0}$
&
 \med{(1),(1),5,\bar{5},24,24}
\\
20
&
     $(H_{u} {\rm i} \tau^{2} \tau^{a} \overline{L^{c}})
     (\tau^{a} H_{d}) ({\rm i} \tau^{2} \tau^{b} H_{u}) 
     (H_{u} {\rm i} \tau^{2} \tau^{b} L)$
&
${\bf 3}^{R}_{0}$, ${\bf 3}^{L}_{0}$,
${\bf 2}^{R}_{-1/2}$, ${\bf 2}^{L}_{-1/2}$,
&
 \med{24,24,5,\bar{5}}
\\
 \hline \rule[0em]{0em}{1.2em}  
  21
&
$(\overline{L^{c}} {\rm i} \tau^{2} \tau^{a} L) 
(H_{u} {\rm i} \tau^{2} \tau^{a})
(\tau^{b} H_{d})
(H_{u} {\rm i} \tau^{2} \tau^{b} H_{u})$
&
${\mathbf 3}^{s}_{+1}$, ${\mathbf 2}^{s}_{+1/2}$ , ${\mathbf 3}^{s}_{+1}$
&
 \med{15,5,15}
 \\
22
&
$(\overline{L^{c}} {\rm i} \tau^{2} \tau^{a} L) 
(H_{d} {\rm i} \tau^{2} \tau^{a})
(\tau^{b} H_{u})
(H_{u} {\rm i} \tau^{2} \tau^{b} H_{u})$
&
${\mathbf 3}^{s}_{+1}$, ${\mathbf 2}^{s}_{+3/2}$, ${\mathbf 3}^{s}_{+1}$
&
 \med{15,40,15}
\\
23
&
$(\overline{L^{c}} {\rm i} \tau^{2} \vec{\tau} L) 
(H_{u} {\rm i} \tau^{2} \vec{\tau})
(H_{u})
(H_{d} {\rm i} \tau^{2} H_{u})$
&
${\mathbf 3}^{s}_{+1}$, ${\mathbf 2}^{s}_{+1/2}$, ${\mathbf 1}^{s}_{0}$
&
 \med{15,5,1}
\\
24
&
$(\overline{L^{c}} {\rm i} \tau^{2} \tau^{a} L) 
(H_{u} {\rm i} \tau^{2} \tau^{a})
(\tau^{b} H_{u})
(H_{d} {\rm i} \tau^{2} \tau^{b} H_{u})$
&
${\mathbf 3}^{s}_{+1}$, ${\mathbf 2}^{s}_{+1/2}$, ${\mathbf 3}^{s}_{0}$
&
 \med{15,5,24}
\\
25
&
$(H_{d} {\rm i} \tau^{2} H_{u}) 
     (\overline{L^{c}} {\rm i} \tau^{2})
     (\vec{\tau} L)
     (H_{u} {\rm i} \tau^{2} \vec{\tau} H_{u})$
&
${\bf 1}^{s}_{0}$, 
${\bf 2}^{L}_{+1/2}$, ${\bf 2}^{R}_{+1/2}$,
${\bf 3}^{s}_{+1}$
&
 \med{1,5,\bar{5},15}
\\
\hline \rule[0em]{0em}{1.2em}  
26
&
$(H_{d} {\rm i} \tau^{2} \tau^{a} H_{u}) 
     (\overline{L^{c}} {\rm i} \tau^{2} \tau^{a})
     (\tau^{b} L)
     (H_{u} {\rm i} \tau^{2} \tau^{b} H_{u})$
&
${\bf 3}^{s}_{0}$, 
${\bf 2}^{L}_{+1/2}$, ${\bf 2}^{R}_{+1/2}$,
${\bf 3}^{s}_{+1}$
&
 \med{24,5,\bar{5},15}
\\
27
&
$(H_{u} {\rm i} \tau^{2} \overline{L^{c}})
({\rm i} \tau^{2} H_{d})
(\vec{\tau} L)
(H_{u} {\rm i} \tau^{2} \vec{\tau} H_{u})$
&
${\bf 1}^{R}_{0}$, ${\bf 1}^{L}_{0}$,
${\bf 2}^{R}_{+1/2}$, ${\bf 2}^{L}_{+1/2}$,
${\bf 3}^{s}_{+1}$
&
 \med{1,1,5,\bar{5},15}
\\
28
&
$(H_{u} {\rm i} \tau^{2} \tau^{a} \overline{L^{c}})
({\rm i} \tau^{2} \tau^{a} H_{d})
(\tau^{b} L)
(H_{u} {\rm i} \tau^{2} \tau^{b} H_{u})$
&
${\bf 3}^{R}_{0}$, ${\bf 3}^{L}_{0}$,
${\bf 2}^{R}_{+1/2}$, ${\bf 2}^{L}_{+1/2}$,
${\bf 3}^{s}_{+1}$
&
 \med{24,24,5,\bar{5},15}
\\
29
&
$(H_{u} {\rm i} \tau^{2} \overline{L^{c}})
(L)
({\rm i} \tau^{2} \vec{\tau} H_{d})
(H_{u} {\rm i} \tau^{2} \vec{\tau} H_{u})$
&
${\bf 1}^{R}_{0}$, ${\bf 1}^{L}_{0}$,
${\bf 2}^{s}_{+1/2}$, 
${\bf 3}^{s}_{+1}$
&
 \med{1,1,5,15}
\\
30
&
$(H_{u} {\rm i} \tau^{2} \tau^{a} \overline{L^{c}})
(\tau^{a} L)
({\rm i} \tau^{2} \tau^{b} H_{d})
(H_{u} {\rm i} \tau^{2} \tau^{b} H_{u})$
&
${\bf 3}^{R}_{0}$, ${\bf 3}^{L}_{0}$,
${\bf 2}^{s}_{+1/2}$,
${\bf 3}^{s}_{+1}$
&
 \med{24,24,5,15}
\\
\hline \rule[0em]{0em}{1.2em}  
31
&
$(\overline{L^{c}} {\rm i} \tau^{2} \tau^{a} H_{d})
     ({\rm i} \tau^{2} \tau^{a} H_{u})
     (\tau^{b} L)
     (H_{u} {\rm i} \tau^{2} \tau^{b} H_{u})$
&
${\bf 3}^{L}_{+1}$, ${\bf 3}^{R}_{+1}$,
${\bf 2}^{L}_{+1/2}$, ${\bf 2}^{R}_{+1/2}$,
${\bf 3}^{s}_{+1}$
&
 \med{15,\overline{15},5,\bar{5},15}
\\
32
&
$(\overline{L^{c}} {\rm i} \tau^{2} \tau^{a} H_{d})
     (\tau^{a} L)
     ({\rm i} \tau^{2} \tau^{b} H_{u})
     (H_{u} {\rm i} \tau^{2} \tau^{b} H_{u})$
&
${\bf 3}^{L}_{+1}$, ${\bf 3}^{R}_{+1}$,
${\bf 2}^{s}_{+3/2}$,
${\bf 3}^{s}_{+1}$
&
 \med{15,\overline{15},40,15}
\\
33
&
$(\overline{L^{c}} {\rm i} \tau^{2} \vec{\tau} H_{d})
({\rm i} \tau^{2} \vec{\tau} H_{u})
(H_{u})
(H_{u} {\rm i} \tau^{2} L)$
&
${\bf 3}^{L}_{+1}$, ${\bf 3}^{R}_{+1}$,
${\bf 2}^{L}_{+1/2}$, ${\bf 2}^{R}_{+1/2}$,
${\bf 1}^{L}_{0}$, ${\bf 1}^{R}_{0}$
&
 \med{15,\overline{15},5,\bar{5},1,1}
\\
34
&
$(\overline{L^{c}} {\rm i} \tau^{2} \tau^{a} H_{d})
({\rm i} \tau^{2} \tau^{a} H_{u})
(\tau^{b} H_{u})
(H_{u} {\rm i} \tau^{2} \tau^{b} L)$
&
${\bf 3}^{L}_{+1}$, ${\bf 3}^{R}_{+1}$,
${\bf 2}^{L}_{+1/2}$, ${\bf 2}^{R}_{+1/2}$,
${\bf 3}^{L}_{0}$, ${\bf 3}^{R}_{0}$
&
 \med{15,\overline{15},5,\bar{5},24,24}
\\
\hline \hline
\end{tabular}
} 
\caption{\label{tab_mediators} Decompositions of the $d=7$ operator $L L H_u H_u H_d H_u$ at tree level. We use the following notation for the mediators: ${\bf X^L_Y}$. The X describes the SU(2) nature, \ie, singlet (1), doublet (2), or triplet (3). The superscript $L$ denotes a left- (L) or right- (R) handed fermion, or a scalar (s). The subscript $Y$ represents the hypercharge $Y \equiv Q - I_3^W$. The SU(5) singlets in parentheses can be already contained in the 24-plets. }
\end{center}
\end{table}

The corresponding $SU(5)$ multiplets are either gauge singlets, 5-plets,
15-plets, 24-plets or 40-plets. The corresponding particles of the complete
multiplet have to have
masses at the TeV scale as not to destroy gauge coupling unification but
only modifying the running of the gauge couplings. Using for example 
\Ref~\cite{Slansky:1981yr} one can easily calculate the additional contributions
of these multiplets to the beta functions which we have summarized in 
\Tab~\ref{tab:betacoefficients}. These values have to be compared with
the MSSM beta functions $b_i=(33/5,1,-3)$ for $U(1)_Y$, $SU(2)_L$ and $SU(3)_C$,
respectively. One sees that all but the 5-plets give rather large contributions
and one might wonder if such large contributions are consistent with
the requirement of perturbativity up to the GUT scale.

\begin{table}[t] \begin{center}
\begin{tabular}{c||cccc}
Multiplet & 5 & 15 & 24 & 40 \\ \hline
$\Delta b_i$ & 1/2 & 7/2 & 5 & 11  \\ 
\end{tabular}
\end{center}
\caption{\label{tab:betacoefficients} Contributions of the various
$SU(5)$ multiplets to the MSSM beta functions.}
\end{table}

At the one-loop level the RGE of the gauge coupling have the well-known
solution
\begin{equation}
\alpha_G = \frac{\alpha_i(Q)}
           {1 - \frac{b_i}{4 \pi} \alpha_i(Q) \ln \frac{M^2_G}{Q^2}} \,,
\label{eq:alphaG}
\end{equation}
where the $Q$ is the scale at which the additional particles are included,
$M_G \simeq 2\cdot 10^{16}$~GeV is the GUT scale and $\alpha_G$ the
value of the gauge couplings at $M_G$. Taking only the MSSM one gets
$\alpha_G \simeq 1/25$. Adding a contribution $\Delta b_i$ to $b_i$ in
\eq~(\ref{eq:alphaG}) we find $\Delta b_i=5$ already implies
that $\alpha_G=1$, see also \cite{Kopp:2009xt} for a related discussion.
As the $5$-plet and the $15$-plet are complex representations, they have
to come always paired with a $\bar{5}$ and $\overline{15}$, respectively,
as otherwise one cannot put a gauge invariant mass term in the superpotential.
Inspecting \Tab~\ref{tab:betacoefficients} we thus conclude that we
can add at most five $5$, $\bar{5}$ pairs or one 24-plet if one is 
willing to accept $\alpha_G=1$. If one would require $\alpha_G \le 1/2$ then
one is restricted to at most four $5$, $\bar{5}$ pairs. In practice the
number is smaller as two-loop effects increase the value of $\alpha_G$
\cite{Esteves:2010ff}. As a consequence, from the possible operators
in \Tab~\ref{tab_mediators} only operators \#1, \#9, \#13 and \#17 are consistent
with perturbativity up to the GUT scale. We note for completeness that in
case of operator \#9 one has to include a corresponding $\bar{5}$ as otherwise
no mass term can be added to the MSSM superpotential. Also for operators
\#9 and \#13  the discussion as for operator \#17 in the previous sections
can be repeated resulting in essentially the same findings.  Note that decompositions \#1,
\#9, and \#13 contain a singlet scalar in the decomposition. Looking at the according superpotentials
\begin{subequations}
 \begin{align}
\begin{split}
W_\text{\#1} &=  \bar{5}_M  10\,  H_{\bar 5} +  10\,  10\,  H_5 +  S\,  S\,  S\, +  S\,  H_5  H_{\bar 5} +  H_5  \bar{5}_M  N\, +  H_5  H_{\bar 5}  \phi\, + \\
&\qquad  N'  N'  \phi\, +  N\,  N'  S\, +  \phi\,  \phi\,  S\, 
\end{split} \\
\begin{split}
W_\text{\#3} &=  \bar{5}_M  10\,  H_{\bar 5} +  10\,  10\,  H_5 +  S\,  S\,  S\, +  S\,  H_5  H_{\bar 5} +  H_5  \bar{5}_M  24_a +  H_5  H_{\bar 5}  \phi\, + \\
&\qquad  24_b  24_b  \phi\, +  24_a  24_b  S\, +  \phi\,  \phi\,  S\, 
\end{split}\\
\begin{split}
W_\text{\#13} &=  \bar{5}_M  10\,  H_{\bar 5} +  10\,  10\,  H_5 +  S\,  S\,  S\, +  S\,  H_5  H_{\bar 5} +  \bar{5}_M  H_5  N\, +  N'  \bar{5}_M  5_s + \\
&\qquad  \bar{5}_s  H_5  \phi\, +  H_5  H_{\bar 5}  \phi\, +  N\,  N'  S\, +  \bar{5}_s  5_s  S\, +  \phi\,  \phi\,  S\, 
\end{split}\,,
\end{align}
\end{subequations}
where the $\phi$ are the scalar singlets and for the other fields we use the notation introduced in the previous sections. Note, that we only give the superfields
but not the details of the couplings. All three superpotentials 
contain the term $H_5  H_{\bar 5}  \phi$, which eventually induces a vev
for the scalar component of  $\phi$. In these cases, a lower than 
$d=7$ dimensional operator will dominate neutrino
mass. This means that our example in  \figu{example} is the only tree level decomposition
which is consistent with
perturbativity up to the GUT scale and neutrino mass generation by a leading $d=7$ 
contribution.

\section{Summary and conclusions}
\label{sec:conclusions} 

We have discussed neutrino mass generation from a $d=7$ effective operator, while the $d=5$ Weinberg operator is suppressed due to a discrete symmetry. The additional suppression by $v^2/\Lambda_{\text{NP}}^2$ compared to the Weinberg operator reduces the new physics scale $\Lambda_{\text{NP}}$ required to describe the smallness of neutrino mass, potentially down to the TeV scale, which makes the mediators accessible at the LHC. We have especially focused on TeV completions in the MSSM and NMSSM frameworks. In order to achieve gauge coupling unification, we have embedded such models in a GUT scenario with an SU(5) subgroup. 

We have chosen one specific example for the decomposition of the $d=7$ operator in \equ{hh7}, see  \figu{example}, which has been discussed earlier in the context of its LHC phenomenology~\cite{Krauss:2011ur}.
It involves two fermionic SU(2) singlets, and two fermionic SU(2) doublets. 
In this case, not only the mediators may be probed at the LHC, but also sensitivity to lepton flavor and perhaps even lepton number violation can be achieved.
 In this study, we have demonstrated that this example is peculiar for two reasons: 1) it can be completed in an SU(5) and described perturbatively up to the GUT scale, and 2) it does not contain any singlet scalars which take a vev and thereby generate neutrino mass by a lower dimensional operator. It is, in fact, the only example which satisfies these two requirements. 

The embedding of the fermionic doublets into SU(2) multiplets implies that additional heavy $d$-quarks are to be introduced, which are strongly constrained by early universe cosmology, such as big bang nucleosynthesis. Therefore, fast enough decay mechanisms for these  $d$-quarks need to be identified. The decay via leptoquarks or higher dimensional operators, which may be generated from physics beyond the GUT scale, is possible, but nevertheless too slow to satisfy the current constraints. The possibility in agreement with cosmological bounds is a decay via a small (discrete) symmetry breaking term, which could be large enough to allow for a sufficient decay rate.

For the specific case of the NMSSM, we have demonstrated that the NMSSM scalar vev $\langle S \rangle$  breaks the discrete symmetry spontaneously, which will lead to a (dominating) neutrino mass term from the Weinberg operator. A simple way out has been shown to introduce an additional scalar and symmetry breaking terms, which can be small enough such that the $d=5$ contribution to neutrino mass is suppressed. We have also shown that these small symmetry breaking terms may describe the origin of deviations from tri-bimaximal mixing, and this may be the origin of the non-vanishing magnitude of $\theta_{13}$.

In conclusion, we have discussed a model which described neutrino mass from physics at the TeV scale, and which is potentially LHC-testable. We have embedded the model in an SU(5) SUSY-GUT, which means that we have confirmed gauge coupling unification and perturbativity up to the GUT scale. Finally, we have established that this model requires the fast enough decay of heavy $d$-quarks, which are needed for the GUT completion of the theory. Therefore, interesting constraints will not only be obtained from LHC, but also from early universe cosmology.

\section*{Acknowledgments}

M.B.K.~likes to thank the physics department of Universit\`a Roma Tre for the hospitality during his stay in Rome. 
This work has been supported by the DFG research training group GRK 1147. 
W.P.~has been supported by the DFG project number PO1337/2-1. W.W.~ would like to acknowledge support from DFG grants WI 2639/3-1 and  WI 2639/4-1.   
This work has also been supported by the FP7 Invisibles network (Marie Curie
Actions, PITN-GA-2011-289442). D.M. acknowledges MIUR (Italy) for financial support under the program 
"Futuro in Ricerca 2010 (RBFR10O36O)".

\newpage



\end{document}